\documentstyle[11pt,IAUS212,twoside,epsf]{article}

\def\edcomment#1{\iffalse\marginpar{\raggedright\sl#1\/}\else\relax\fi}
\reversemarginpar

\begin{document}
\vspace*{1cm}
\title{Dispersal of massive star products and consequences for
galactic chemical evolution}
 \author{M. S. Oey}
\affil{Lowell Observatory, 1400 W. Mars Hill Rd., Flagstaff, 
	AZ\ \ \ 86001, USA}

\begin{abstract}
The processes that disperse the products of massive stars from their
birth sites play a fundamental role in determining 
the observed abundances.  I discuss parameterizations for
element dispersal and their roles in chemical evolution, with an
emphasis on understanding present-day dispersion and homogeneity in
metallicity.  Turbulence dominates mixing processes, with
characteristic timescales of order $10^8$ yr, implying significant
dilution of metals into the ISM.  This permits a rough estimate of the 
metallicity distribution function of enrichment events.  Many systems,
including the Milky Way and nearby galaxies, show metallicity
dispersions that as yet appear consistent with pure inhomogeneous evolution. 
There are also systems like I~Zw~18 that show strong homogenization,
perhaps tied to small galaxy size, high star formation rate, and/or
superwinds. 
\end{abstract}

\section{Introduction}

The dispersal of the nucleosynthetic products of massive stars plays a
fundamental role in galactic chemical evolution.  Our ability to
decipher the chemical signatures of star formation history depend
upon understanding the dispersal processes.  The growing
database and improving accuracy in elemental abundance determinations
increasingly is revealing the magnitude of abundance dispersions, and
setting limits on uniformity.  It is apparent that the magnitude of these
dispersions results directly from the balance between interstellar
transport processes and the timescale for global star formation.
This field of study is presently still in its infancy, but I will
discuss here some of our rudimentary knowledge.

Dispersal of newly-synthesized elements takes place on three levels:
{\bf 1.} Local {\bf mixing} with the ambient interstellar medium
(ISM); {\bf 2.} Global {\bf homogenization} of a system or subsystem;
{\bf 3.  Outflow 
/ inflow} of metals from the considered system.  The local mixing
processes describe the immediate dispersal of elements associated
with individual star-forming regions into areas of
roughly uniform metallicity.  Homogenization of the system then
describes the mixing of these individual patches toward globally
uniform abundances.  Finally, outflows and inflows consider the
transport of metals beyond the system altogether.  Other
presentations in this volume (e.g., Strickland; Matteucci) consider the
last; here, I will consider only the first two processes, which apply
only within a given system.

\section{The simplest models}

Rudimentary representations of mixing essentially correspond to models
for inhomogeneous chemical evolution.  In the very simplest model, the
metallicity of a system is considered to be built up by identical
individual patches of contamination representing, for example,
individual supernova remnants (SNRs).  If these patches are all
uniform in size and metallicity, then, as pointed out by Edmunds
(1975), the expected present-day 
dispersion in metallicity $Z$ is $Z_1/\sqrt{n}$, where
$Z_1$ is the present-day metallicity and $n$ is the number of SNRs
contributing to $Z_1$ at any given point.  For the solar neighborhood,
this predicts a present fractional dispersion in metallicity of
roughly $10^{-3}$, which is orders of magnitude more uniform than the
observed dispersion that is of order $10^{-1}$.

Individual regions of contamination are not uniformly identical,
however.  Thus, in constructing a quantitative model for
Simple Inhomogeneous Evolution (SIM), we adopted a metallicity
distribution function (MDF) $f(Z)$ for the unit patches that build up a
system's metallicity (Oey 2000).  Instead of individual SNRs, the
products from core-collapse supernovae (SNe) are assumed to be distributed
via superbubbles resulting from the spatial correlation of SNe in OB
associations.  Oey \& Clarke (1997) showed that most ordinary
superbubbles at some stage become pressure-confined, resulting in
final radii $R$ that are related to the input mechanical luminosity
$L$ as $R\propto L^{1/2}$; thus the total superbubble volume
$V \propto L^{3/2}$.  The total mass of new metals produced by these
same SNe, however, is directly proportional to $L$.
We therefore find that
\begin{equation}\label{ZvsR}
Z\propto L^{-1/2} \propto R^{-1}
\end{equation}
for any given superbubble, implying that the larger objects yield lower
metallicity because of the dilution into greater volume.
Based on the empirical {\sc H~ii} region luminosity function and OB
association mass function, Oey \& Clarke (1997) derived a size
distribution for superbubbles $N(R)\propto R^{-3}$.  Together with
equation~1,
we therefore have
\begin{equation}
f(Z)\propto Z^{-2}
\end{equation}
for the parent MDF of contaminating patches at any given
point in the ISM (Oey 2000). 

Of equal importance as the functional form for $f(Z)$ is its
characteristic mean metallicity $\delta Z$.  If $\delta Z$
is small, then a system requires more generations of star formation at
any given point to build up to a given present-day $Z_1$.  Thus we see
that the relationship between the SN yield and 
interstellar mixing is critical in determining both $\delta Z$ and also
the range of $f(Z)$.  Assuming no mixing beyond the final superbubble
radius, and taking fiducial maximum and minimum superbubble radii of
1300 and 25 pc, yields extreme metallicities corresponding to
[Fe/H]$_{\rm min} = -2.6$ and [Fe/H]$_{\rm max} = -0.6$.  Since metal-poor
stars show metallicities ranging continuously down to [Fe/H]$\sim
-4.0$, this demonstrates that metals must mix in the ISM much farther
beyond the superbubble radius.  Note that throughout this
contribution, [Fe/H], as the principal stellar metallicity indicator,
is used as a surrogate for metallicity resulting from Type~II SNe.
We do not consider here the evolution of products from lower-mass stars.

\section{Interstellar mixing processes}

Thus we are led to consider the physical processes of mixing in the
interstellar medium.  The conventional view of the global ISM has been
that of distinct hot ($10^6$ K), warm ($10^4$ K), and cold ($10^2$ K)
phases in rough pressure equilibrium.  However, recent studies are
instead suggesting that, while these temperature phases do exist, as
they must because of the cooling function, they appear to be less
distinct than has been thought (e.g., V\'azquez-Semadeni 2002;
Kritsuk \& Norman 2002).  The cool clouds, in particular, are
apparently transient, unconfined features.  This new
characterization of the ISM suggests even more strongly that turbulent
processes are likely to dominate kinematics.

Turbulence has already been implicated in the past by the apparent
existence of hierarchical scales for energy injection.  Roy \& Kunth
(1995) argued that mixing timescales for small-scale processes are
consistently less than those for large-scale processes.  They
considered large, intermediate, and small scales respectively driven
by galactic rotation and shear, superbubbles and cloud collisions, and
molecular diffusion.  The observed spatial power spectrum of structure
in the ISM is consistent with such hierarchical energy input, which
plausibly drives an energy cascade resulting in a Kolmogorov
power-law, compatible with the observed spatial power spectrum in the
ISM (e.g., Lazarian 2002; Scalo 1987).  
Thus turbulence has been identified by a number of authors as the
dominant agent for mixing in the ISM (e.g., Roy \& Kunth 1995;
Tenorio-Tagle 1996).

We therefore consider the two primary candidates for mass transport
to be atomic diffusion and turbulent mixing.  While previous studies
have also considered cloud collisions (Bateman \& Larson 1993; Roy \&
Kunth 1995), this process appears less viable for the new paradigm of
transient, unconfined cold clouds.  It also requires the cloud
velocities to be larger than the random velocities of the ISM
associated with the other transport mechanisms.  Thus we defer
consideration of cloud collisions to the mentioned references.
Oey (2002) recalculates the diffusion coefficients for different ISM
temperature phases.  Table~1 confirms, from a more detailed analysis, the
finding by Tenorio-Tagle (1996) that diffusion is a highly inefficient
process for mass transport, even for the hot ionized medium (HIM).

\begin{table}
\caption{Parameters for O diffusion in diffuse ISM}
\smallskip
\begin{center}
\begin{tabular}{cccccc}
Phase & $T$ & $n$(H) & O ion & log $D_{12}$ & $r_{\rm rms}\ ^a$ \\
& (K) & (cm$^{-3}$) & & ($\rm cm^2\ s^{-1}$) & (pc) \\
\tableline
CNM & $1\times 10^2$ & 1.0 & O$^0$ & 18.25 & 0.06 \\
WNM & $8\times 10^3$ & 0.3 & O$^0$ & 20.72 & 1.0 \\
WIM & $1\times 10^4$ & 0.1 & O$^{+2}$ & 18.04 & 0.05 \\
HIM & $1\times 10^6$ & 0.003 & O$^{+5}$ & 23.64 & 30 \\
\end{tabular}
\end{center}
$^a$ {\scriptsize Diffusion length for $1\times 10^8$ yr.}
\end{table}

While turbulent processes are difficult to quantify, we describe here
some crude estimates (see Oey 2002 for details).  Bateman \& Larson
(1993) give an expression for the characteristic mixing distance:
\begin{equation}\label{eqBL}
r_{\rm trb} = \bigl(2/3\ v_{\rm trb}\ l_{\rm trb}\ t\bigr)^{1/2} 
	= 58.3\ \bigl(v_{\rm trb}/{\rm km\ s^{-1}}\ \bigr)^{1/2}\ 
	\rm pc \quad ,
\end{equation}
where $v_{\rm trb}$ and $l_{\rm trb}$ are the turbulent velocity
and associated length scale, respectively; and the coefficient 58.3
results for $l_{\rm trb} = 50$ pc and $t = 1\times 10^8$ yr.  Thus we see,
in comparison with Table~1, that turbulence is orders of magnitude more
efficient than diffusion.

In estimating the mixing time for newly synthesized elements
originating in hot
superbubble interiors, we argue that the most efficient length scale
is that associated with the size of the considered region, namely,
the superbubble radius:  $l_{\rm trb} \sim R$.  To uniformly mix in the
ISM, and especially, to become incorporated in subsequent star
formation, the metals need to cool.  For turbulence, heat
transport and mass transport are analogous processes, therefore
turbulence most likely dominates the cooling process as well.
We assume that we need roughly an order of magnitude more cool
particles than hot particles to cool the hot gas by turbulent heat
transport (and assistance from the cooling curve), thus yielding rough
estimates:
\begin{eqnarray}\label{tcool}
t_{\rm cool,W} & \simeq & 0.15\ \l_{\rm trb}(\rm pc)\ \ Myr\ \ \ (WM) \\
t_{\rm cool,C} & \simeq & 0.97\ \l_{\rm trb}(\rm pc)\ \ Myr\ \ \ (CNM) \ , 
\end{eqnarray}
for $v_{\rm trb} = 16\ \rm km\ s^{-1}$ and 1.6 km s$^{-1}$ for cooling into
warm medium (WM; both neutral and ionized) and cold
neutral medium (CNM), respectively.  These expressions are comparable
to similar analytic estimates by de Avillez \& Mac Low (2002).
Assuming that new metals need at least this cooling time to join the
WM or CNM, equations~4 and 5 suggest
they can mix out to
\begin{equation}\label{rmixturb}
r_{\rm mix}
	= 2^{-1/3} \epsilon_2
	\Bigl(\frac{v_1}{v_2}\Bigr)^{1/2}\ R \ \ \ \rm (WM, HIM) 
\end{equation}
\begin{equation}\label{rmixturbCNM}
r_{\rm mix} = \Bigl[\frac{2}{3} v_2 R\ 
	(\tau_n - t_{\rm cool,C} - t_f)\Bigr]^{1/2} \ \ \ \rm (CNM)
\end{equation}
where subscripts 1 and 2 denote parameters for the dispersing and
ambient media, respectively, and $\epsilon$ is a factor depending on
their density ratio.  The star formation duty cycle timescale is
represented by $\tau_n$ and $t_f$ is the age at which the parent
superbubble attains its final pressure-confined radius.
For a density ratio $n_1/n_2 = 0.1$ and
$v_1/v_2 = 10$, equation~6
reduces to $r_{\rm mix} =
\sqrt{10}\ R$.  Thus, we estimate a characteristic dispersal length
that depends on the balance between the different ISM temperature
phases:
\begin{equation}\label{rmixtot}
r_{\rm mix,tot} = \Bigl[R^3 + f_{\rm C} r_{\rm C}^3 + f_{\rm W} r_{\rm W}^3
	+ f_{\rm H} r_{\rm H}^3 \Bigr]^{1/3} \quad ,
\end{equation}
where $f_{\rm C},\ f_{\rm W},$ and $f_{\rm H}$ are the mass fraction
of cold, warm, and hot ISM, respectively.  The ISM phases are assumed
to be individually continuous, and, for cooling by turbulent mixing to
dominate, requires $f_{\rm C} > f_{\rm W} > f_{\rm H}$.

Figure~1$a$ shows limits for the characteristic dispersal length $r_{\rm
mix,tot}$ as a function of parent superbubble radius $R$ for $\tau_n$ =
500 Myr (dashed lines) and 200 Myr (solid lines), as given by
equations~6 -- 8.
We see that the 
the relation between $R$ and $r_{\rm mix,tot}$ is virtually linear,
with a slope between 0.7 and 3.1.  This suggests we may crudely
estimate that metals synthesized by massive stars are dispersed to
radii up to $\sim 3$ times the original superbubble radius.

\begin{figure}
\plotone{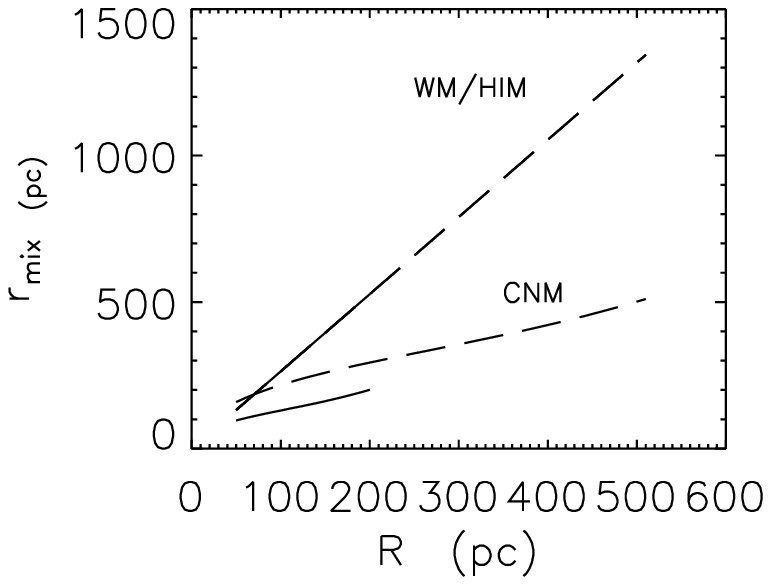}
\plotfiddle{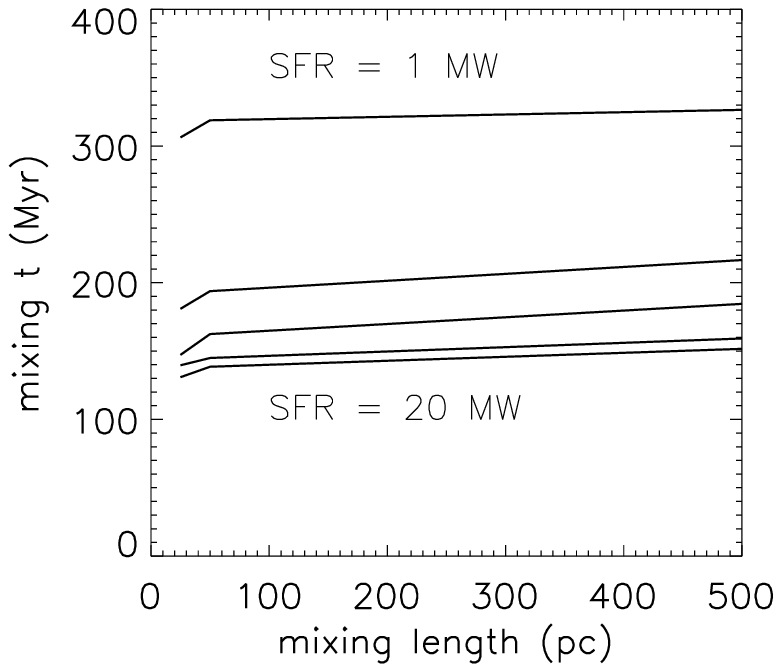}{10pt}{0}{70}{70}{-40}{-360}
\caption{($a$) Limits for dispersal length $r_{\rm mix,tot}$ estimated
from equations~6 -- 8, for $\tau_n = 500$ (dashed) and 200 Myr
(solid), respectively.  ($b$) Mixing time as a function of mixing
length from Table~1 of de Avillez \& Mac Low (2002), for SN rates of
1, 5, 10, 15, and 20 times the Galactic rate.}
\end{figure}

That the metals disperse to large distances while within the HIM has
been suggested by, e.g., Kobulnicky (1999).  Starburst galaxies
show spatial uniformity in nebular abundances to $\la 0.2$ dex
with no evidence of self-enrichment near major star-forming complexes;
NGC~1569 is a notable example where metal production expected from the
supercluster A must be dispersed within a diameter of at least 100 pc to
explain the observed abundance uniformity (Kobulnicky \& Skillman
1997).  One of the only existing candidate counter-examples are provocative
observations by Cunha \& Lambert (1994) that suggest a metal-enriched
sub-group of stars within the Orion Nebula. 

In addition, actual hydrodynamical simulations of turbulent
mixing processes are finally beginning to be realized.  De Avillez \&
Mac Low (2002) carry out such a study using a 3D, PPM, 3-level
adaptive mesh refinement code.  They assume a multi-temperature,
density-stratified ISM with turbulence driven exclusively by SNe.
Intriguingly, they find that the mixing time in their simulations is
essentially independent of the mixing length.  Figure~1$b$ plots the
mixing time as a function of mixing length from their Table~1, for
different SN rates.  This surprising result, which contradicts
conventional mixing length theory, clearly bears further investigation.
More intuitively, they find that the mixing time is
more sensitive to the input SN rate (see Figure~1$b$).  Note that for
SN rates of~1 -- 10 times the Galactic rate, the mixing timescales are
of order a few times $10^8$ yr, as we estmate above.

\section{Consequences for chemical evolution}

With a coarse idea for mixing timescales and corresponding dispersal
lengths, we now revise the estimates for the MDF $f(Z)$ of the
parent metallicities that drive galactic chemical evolution.
Earlier above, we used the superbubble radii themselves as
characteristic dispersal lengths.  In that case, the fiducial maximum
and minimum radii of 1300 and 25 pc correspond to the unrealistically
high values of minimum and maximum [Fe/H] of --2.6 and --0.6,
respectively.  These estimates can now be reduced by roughly an order
of magnitude, since we found above that new metals are dispersed into
a volume roughly an order of magnitude greater during their cooling
time.  Thus we obtain a range for the parent $f(Z)$ corresponding to
$-3.6 \la \rm [Fe/H] \la -1.9$.  This implies a characteristic mean
value for the typical enrichment unit $\delta Z\ \rm of\ [Fe/H] \sim
-3.0$.  Argast et al. (2000) use a similar stochastic approach for
their simulation of chemical evolution based on contamination by
individual SNRs; we note that they adopt $\delta Z\ \rm of\ [Fe/H]
\sim -2.8$.  Thus, there is some reassurance that these independent
models assume similar values.

Note that the lowest observed metallicities may offer an important
constraint on the minimum values of the parent MDF.  For a closed-box
model, Audouze \& Silk (1995) emphasize the existence of a
low-metallicity threshold $Z_{\rm min}$ for $f(Z)$:
in a system with no global homogenization, there can be no regions
with metallicities between zero and $Z_{\rm min}$ of the enrichment
units.  It has been suggested that the currently observed
low-metallicity limit of [Fe/H] $\sim -4.0$ for Galactic halo stars
(Beers 1999) represents such a real threshold.  If so, then our
crudely estimated [Fe/H]$_{\rm min}$ is reasonably consistent with
this value.  Additional uncertainty in our estimate results from the
assumed SN yield of 10 M$_\odot$ of metals per SN; this value could be
reduced by up to an order of magnitude, further improving the rough
estimate with observations.  It is thus of great interest to determine
whether the empirical threshold indeed exists.

The enrichment units represented by $f(Z)$ can now be used to
drive models for inhomogeneous chemical evolution.  As described
above, we argued that $f(Z)\propto Z^{-2}$ within the
$Z_{\rm min}$ and $Z_{\rm max}$ modified above.  The Simple
Inhomogeneous Evolution model of Oey (2000) can then be used to
predict the expected dispersion in present-day metallicity resulting
from pure inhomogeneous evolution.  Note that in the presence of
homogenization, the system would then tend toward description by the
homogeneous Simple Model.

Figure~2 shows predicted instantaneous MDFs for two systems, assuming
the parent $f(Z)$ estimate above:  $f(Z)\propto Z^{-2}\ ,\ -3.6 < \rm
[Fe/H] < -1.6$.  For the model evolved to $\sim 0.1 Z_\odot$, we find
a predicted [O/H] dispersion of 0.28 dex, and for the model evolved to
$\sim Z_\odot$, the predicted dispersion is 0.13 dex.  (See Oey 2002
for the conversion between [O/H] to [Fe/H].)  Observations of the
metallicity dispersion in the solar neighborhood do appear to be
in reasonable agreement:  new abundances of 70 B stars by
Daflon \& Cunha (this volume) show a scatter of around 0.2 dex, and
measurements from lower-mass field stars and open clusters by Twarog
et al. (1997) show a dispersion of around 0.1 dex around the solar
circle.  The age-metallicity relation in the solar neighborhood 
shows a dispersion of order 0.2 dex for the youngest F and G stars
(Edvardsson et al. 1993).  

\begin{figure}
\plotone{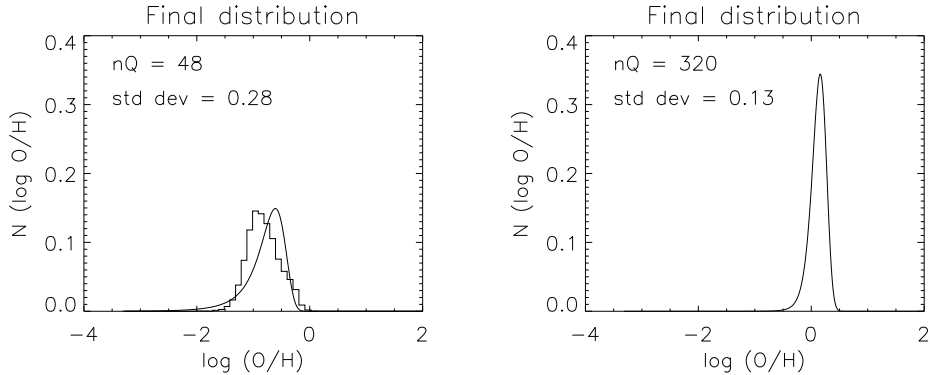}
\caption{Instantaneous metallicity distribution functions predicted by 
a SIM model for systems at roughly 0.1 $Z_\odot$ and 1.0 $Z_\odot$.
These correspond to predicted standard deviations in metallicity
dispersion of 0.28 dex and 0.13 dex, respectively.
The model assumes a characteristic enrichment event $\delta Z$ of 
[O/H] $\sim -2.4$ (see text for details).}
\end{figure}

The interstellar deuterium abundance
deserves special mention:  since the evolution of D/H is driven 
almost exclusively by depletion from the cosmological abundance by
star formation, predictions for present-day D/H are independent of any
stellar yields.  This different evolution process for D/H therefore
will offer an important independent constraint on chemical evolution (see,
e.g., Clayton 1985; Tosi et al. 1998; de Avillez \& Mac Low 2002), although
to date no quantitative model exists for interpreting the dispersion in D/H.
Recent observations of the local D/H abundance from {\sl FUSE} and
{\sl IMAPS} are revealing a dispersion of $\sim 0.15$ dex at distances
of a few 100 pc from the sun (e.g., Moos et al. 2002; Sonneborn
et al. 2000; Jenkins et al. 1999).  Within the Local Bubble, D/H is
significantly more uniform, as expected for efficient mixing within
its hot gas.

Beyond the Milky Way, observations of nearby dwarf starburst galaxies
like NGC 4214, NGC 1569, and NGC 5253 (Kobulnicky \& Skillman 1996,
1997; Devost et al. 1997) have been found to show nebular abundance
dispersions of $\la 0.2$ dex.  Similar dispersions are found from 
A and F supergiant abundances in Local Group dwarfs like NGC 6822 (see
Venn, this volume).  It has been suggested that the abundance
dispersions are unexpectedly uniform, but as shown above, our Simple
Inhomogeneous Models estimate metallicity dispersions of
roughly 0.1 -- 0.3 dex in the range 0.1 -- 1 $Z_\odot$ that are
represented by these galaxies.  For the current empirical accuracy and
large model uncertainties, the comparison between observations and
predictions still looks good.  There is, however, lack of
evidence for self-enrichment, as noted in \S 2 above, which is
consistent with the extended cooling times of order $10^8$ Myr
determined above.  It is somewhat unclear whether the currently
reported dispersions are close to real values or represent 
upper limits, and it is thus vital to determine them more accurately.

An important feature of inhomogeneous chemical evolution is the
expectation of increasing scatter at low metallicities:  the
dispersion is naturally larger when stochastic enrichment
dominates, as it does in the earliest stages when $Z \sim \delta Z$.
(Audouze \& Silk 1995).  Such an increase in stellar abundance
dispersions is well-known to exist, especially for abundance ratios,
at [Fe/H] $\la -1.5$ (e.g., McWilliam et al. 1995; Ryan et al. 1996).
Stochastic simulations by Argast et al. (2000) indeed reproduce many
features of this pattern in  the dispersion, and as is evident above,
our SIM models also show this trend.  However, 
observational discrepancies exist, in particular the extremely
metal-poor dwarf galaxy I~Zw~18.  While the above stellar dispersions
are $\ga$ 0.5 dex, the nebular abundances of [O/H] show a typical
scatter of $\la 0.15$ dex (V\'\i lchez \& Iglesias-P\'aramo 1998;
Legrand et al. 2000).  Thus the present-day metallicity of I~Zw~18 is
significantly more uniform than expected from our purely inhomogeneous
model.  This suggests that the system homogenizes on a
timescale faster than the star formation duty cycle.  Such
homogenization could be facilitated by different factors:  a) The
galaxy's small size ($\sim 1$ kpc) shortens both the homogenization length
and time scale; b) a high star formation rate would shorten the
homogenization time scale (de Avillez \& Mac Low 2002);
and c) the small gravitational potential could allow the
bulk of new metals to be ejected from the galaxy, lengthening the
effective star formation duty cycle.  Hence, understanding the
transport of metals in this and similar galaxies offers vital leverage
on the mixing and homogenization processes in general.

\section{Conclusion}

In summary, the dispersal of newly synthesized elements takes place on
at least three scales:  localized mixing, global homogenization, and
outflow/inflow from the system.  It has emerged that the transport
mechanisms are dominated by turbulence.  The dispersal length scale
for localized mixing 
determines the dilution of new metals in the ISM, and therefore sets
the MDF of the parent enrichment units $f(Z)$ that drive chemical
evolution.  We crudely estimate that, depending on the balance of 
ISM temperature phases, the hot, metal-bearing gas will have cooling
times of order $10^8$ yr and disperse to roughly a decade in distance
beyond the original radius of the superbubble created by the SNe.
This is consistent with analytic and numerical hydrodynamical
simulations of mixing by de Avillez \& Mac Low (2002).  An important
constraint on the minimum 
values for $f(Z)$ should be given by the observed low-metallicity
threshold; we find that the current observed limit of [Fe/H]$\sim -4$
for halo stars (Beers 1999) is compatible with our crude,
order-of-magnitude estimate.

This rough understanding of element dispersal and corresponding
estimate for the $f(Z)$ of individual enrichment units can then be
incorporated into Simple Inhomogeneous Models for chemical evolution.
These yield estimates for the present-day dispersion in metallicity,
which may be compared with observations.  We estimate dispersions of
0.1 -- 0.3 dex in the metallicity range $0.1 - 1.0 Z_\odot$, which is
thus far consistent with data for the solar neighborhood, Local Group
galaxies, and several nearby starburst galaxies, thereby suggesting
that global homogenization is thus far unnecessary to explain
observations.  However, the scatter in abundance is expected to
increase at lower metallicities.  While this is seen in Milky Way
stellar abundances, the extremely metal-poor galaxy I~Zw~18 shows
metallicities that are more uniform 
than expected from the SIM.  This implies that this galaxy has a
faster homogenization time relative to its star formation duty cycle,
perhaps caused by its small size, high star formation rate, and/or
outflow of hot, metal-bearing gas.  It is evident that our
understanding of these processes are still rudimentary, and more data
and modeling are necessary.

\acknowledgments

I am indebted to the conference organizers for inviting this
presentation, and for partial support to attend this Symposium.  
I am also grateful to Miguel de Avillez for helpful communications.

\end{document}